# Ionization of Hydrogen by Electron Vortex Beam

A. L. Harris, A. Plumadore, and Z. Smozhanyk

Physics Department, Illinois State University, Normal, IL, USA 61790

**Abstract**

Optical vortex beams have an extensive history in terms of both theory and experiment, but only recently have electron vortex beams been proposed and realized. The possible applications of these matter vortex waves are numerous, but a fundamental understanding of their interactions with atoms and molecules has not yet been developed. In this work, fully differential cross sections for fast (e,2e) collisions using electron vortex projectiles with small amounts of quantized orbital angular momentum are presented. A comparison is made with the fully differential cross sections using plane wave projectiles and a detailed study of angular momentum transfer is included. Results indicate that ionization by electron vortex beam projectiles is much less likely than for plane wave projectiles, and the angular momentum of the incident electron is transferred directly to the ionized electron.

## 1. Introduction

Vortex beams are freely propagating beams characterized by their non-zero orbital angular momentum (OAM) around the propagation direction and phase singularity at the center of the vortex. Their quantized topological structure with spiraling wave fronts has been widely studied in optical contexts beginning with Nye and Berry in 1974 [1]. In the optical case, vortex beams are used extensively in applications such as optical tweezers [2,3], microscopy [4,5], micromanipulation [6], astronomy [7], and many others. In these applications, the transfer of OAM allows for the control and manipulation of atoms and molecules on the nanoscale. Only recently it was demonstrated that similar vortex beams could be generated with electrons [8]. Since then, several experimental groups have produced electron vortex beams (EVBs) using various methods [9-12], but applications of EVBs are still in the speculative state.

Clearly, the possibilities for EVBs are numerous, and all involve their interaction with matter at the atomic scale. Unfortunately, very little is known about how EVBs interact with individual atoms, and because the experimental generation of EVBs has only recently become possible, there are no experimental results yet for collisions between EVBs and individual atoms. Also, very little theoretical work exists on this topic, with only a handful of theoretical studies to date for EVB collisions with hydrogen atoms [13-17]. If EVBs are to be used for any of the above applications, it is crucial to understand how electrons with non-zero orbital angular momentum (OAM) interact with atoms on a fundamental level.

To date, most of the theoretical work for collisions between EVBs and individual atoms has been performed by the group of van Boxem, Partoens, and Verbeeck [14,15], in which they used the First Born Approximation (FBA) to study potential scattering and inelastic excitation collisions. In [14], Rutherford potential scattering was examined for incident Bessel beams scattering from a Yukawa potential, and it was shown that for non-zero angular momentum, the cross section along the beam direction was zero, indicating a vortex wave is emitted from the scattering center. In a follow-up study, the group expanded upon the Rutherford scattering model to look at excitation of hydrogen by EVB projectiles [15], and a set of selection rules was derived showing that OAM is transferred directly from the projectile to the atom. Several other studies have also examined OAM transfer to atoms, molecules, and thin films [18-21].

The two studies discussed above represent the majority of the theoretical work that has been performed for collisions involving EVB projectiles. Clearly this leaves large gaps in our understanding of the interaction of EVBs with even simple atomic targets. Ionizing collisions using EVB projectiles represent a much broader field of study due to the additional degrees of freedom offered by a second free electron after the collision. Here, we study ionization of atomic hydrogen using EVB projectiles and specifically explore angular momentum transfer to the ionized electron. We show that the theoretical approach is easily generalized to more complex targets and more sophisticated models than the FBA. Atomic units are used throughout unless otherwise noted.

## 2. Theory

Because very little work exists for collisions between EVBs and atoms, we present here fully differential cross sections (FDCS) using the first Born approximation (FBA) for ionization of hydrogen. In an FBA model, the free electrons are modeled as plane waves, however, in the case of EVBs, the incident projectile is a vortex beam, such as a Bessel beam. In a perturbative model such as the FBA, the FDCS is proportional to the square of the transition matrix $T_{fi}$

$$\frac{d^3\sigma}{d\Omega_1 d\Omega_2 dE_2} = \mu_{pa}^2 \mu_{ie} \frac{k_f k_e}{k_i} |T_{fi}|^2 \tag{1}$$

with

$$T_{fi} = <\Psi_f|V_i|\Psi_i>. \tag{2}$$

Here $\mu_{pa}$ is the reduced mass of the projectile and target atom, $\mu_{ie}$ is the reduced mass of the proton and the ionized electron, $\vec{k}_f$ is the momentum of the scattered projectile, $\vec{k}_e$ is the momentum of the ionized electron, and $\vec{k}_i$ is the momentum of the incident projectile. Insertion of complete sets of position states into Eq. (2) results in an integral over all position space for each of the particles in the collision. For EVBs, it is most convenient to use cylindrical coordinates for the projectile and spherical coordinates for the atomic electron. Then, the momenta can be written in terms of their respective components as $\vec{k}_i = k_{i\perp}\hat{\rho}_{1i} + k_{iz}\hat{z}_1$, $\vec{k}_f = k_{f\perp}\hat{\rho}_{1f} + k_{fz}\hat{z}_1$, and $\vec{k}_e = k_e\hat{r}_2$. We consider here what is traditionally referred to as coplanar geometry, in which the final projectile and ionized electron momentum lie in the same plane.

The initial state wave function is a product of the incident projectile Bessel beam wave function $\chi_{\vec{k}_i}(\vec{r}_1)$ and the target hydrogen atom wave function $\Phi(\vec{r}_2)$

$$\Psi_i = \chi_{\vec{k}_i}(\vec{r}_1)\Phi(\vec{r}_2). \tag{3}$$

We consider only the case in which the vortex beam is centered on the atom, such that the beam's orbital angular momentum axis intersects the center of the atom. The Bessel beam is the free particle solution to the Schrödinger equation in cylindrical coordinates and is given by

$$\chi_{\vec{k}_i}(\vec{r}_1) = \frac{e^{il\phi_{ki}}}{2\pi} J_l(k_{i\perp}\rho_1) e^{ik_{iz}z_1}, \tag{4}$$

where $\phi_{ki}$ is the azimuthal coordinate for the incident projectile momentum, $l$ is the quantized orbital angular momentum of the incident projectile, and $J_l(k_{i\perp}\rho_1)$ is the Bessel function. The longitudinal and transverse projectile momentum appear explicitly in this equation as $k_{iz}$ and $k_{i\perp}$, as well as the quantized OAM $l$. This Bessel beam can be rewritten as a superposition of plane waves [15] such that

$$\chi_{\vec{k}_i}(\vec{r}_1) = \frac{(-i)^l}{(2\pi)} \int_0^{2\pi} d\phi_{ki}\, e^{il\phi_{ki}} e^{i\vec{k}_i \cdot \vec{r}_1}. \tag{5}$$

The final state wave function is a product of the scattered projectile wave function $\chi_{\vec{k}_f}(\vec{r}_1)$ and the ionized electron wave function $\chi_{\vec{k}_e}(\vec{r}_2)$

$$\Psi_f = \chi_{\vec{k}_f}(\vec{r}_1)\chi_{\vec{k}_e}(\vec{r}_2). \tag{6}$$

The perturbation $V_i$ is the Coulomb interaction between the projectile and target atom, which is given by

$$V_i = \frac{-1}{r_1} + \frac{1}{r_{12}} \tag{7}$$

for an electron incident on a hydrogen atom.

As in [15], we assume that the scattered projectile leaves the collision as a plane wave given by

$$\chi_{\vec{k}_f}(\vec{r}_1) = \frac{e^{i\vec{k}_f \cdot \vec{r}_1}}{(2\pi)^{3/2}}. \tag{8}$$

This allows us to write the vortex transition amplitude in terms of the plane wave transition amplitude, which is easily calculated and well-known. Combining equations (3)-(8) yields the following expression for the vortex beam transition amplitude [15]

$$T_{fi}^V = \frac{(-i)^l}{(2\pi)} \int_0^{2\pi} d\phi_{ki} e^{il\phi_{ki}} T_{fi}^{PW}(q), \tag{9}$$

where $T_{fi}^{PW}(q)$ is the transition amplitude for an incident plane wave scattering on a hydrogen atom. A major advantage of Eq. (9) is that because the vortex beam transition amplitude is written in terms of the plane wave transition amplitude, it is easily generalized to more complex targets or more sophisticated models. While the derivation of Eq. (9) requires that the scattered projectile be a plane wave, it has no such requirement for the treatment of the ionized electron or the target atom. Thus, one can calculate $T_{fi}^{PW}$ using any theoretical technique or any target atom and simply insert the result into Eq. (9) to find the EVB amplitude. One only needs to take care of the calculation of the "momentum transfer" $\vec{q} = \vec{k}_i - \vec{k}_f$, which in the case of an incident vortex beam must be written in terms of its parallel and perpendicular components. Keeping close track of the azimuthal angle for each of the momenta, the magnitude of the momentum transfer can be written as

$$q^2 = k_i^2 + k_f^2 - 2k_{iz}k_{fz} - 2k_{i\perp}k_{f\perp}\cos(\phi_{ki} - \phi_{kf}), \tag{10}$$

where $\phi_{ki} = 0$ for the coplanar geometry used here.

This expression for momentum transfer combined with the Bessel beam expression as a superposition of plane waves makes it clear that there is no longer a single momentum transfer vector because of the dependence on $\phi_{ki}$ [15]. In the case of either excitation of hydrogen or ionization of hydrogen where the ionized electron is treated as plane wave, evaluation of $T_{fi}^{PW}$ can be performed analytically. Then, Eq. (9) is evaluated numerically. Note that if the orbital angular momentum of the incident projectile is zero and its momentum vector is oriented along the z-axis, the momentum transfer reduces to the standard form and evaluation of Eq. (9) results in the appropriate plane wave transition amplitude.

*2.1 Ionization from ground state*

For ionization from the ground state, the plane wave transition amplitude is

$$T_{fi}^{PW} = \frac{1}{(2\pi)^3} \int d\vec{r}_1 d\vec{r}_2 e^{i\vec{q}\cdot\vec{r}_1} \frac{e^{-i\vec{k}_e\cdot\vec{r}_2}}{(2\pi)^{3/2}} V_i \Phi(\vec{r}_2), \tag{11}$$

where the ground state hydrogen wave function is given by

$$\Phi_{1s}(\vec{r}_2) = \frac{e^{-r_2}}{\sqrt{\pi}}. \tag{12}$$

Evaluation of the integral $\vec{r}_1$ in Eq. (11) yields

$$T_{fi}^{PW} = \frac{4\pi}{(2\pi)^3 q^2} \int \frac{e^{-i\vec{k}_e\cdot\vec{r}_2}}{(2\pi)^{\frac{3}{2}}} \left(-1 + e^{i\vec{q}\cdot\vec{r}_2}\right) \Phi(\vec{r}_2) d\vec{r}_2. \tag{13}$$

From here, the remaining integral over $\vec{r}_2$ can be evaluated either numerically or analytically. Numerical evaluation of Eq. (13) provides easy flexibility in the choice of target wave function or ionized electron wave function. One could easily replace these with an excited state of hydrogen or a more sophisticated treatment of the ionized electron. For ionization from the ground state, analytical evaluation of Eq. (13) yields

$$T_{fi}^{PW-1s} = \frac{\sqrt{2}}{\pi^3 q^2} \left[\frac{1}{(1+k_e^2)^2} - \frac{1}{\left(1+|\vec{q}-\vec{k}_e|^2\right)^2}\right], \tag{14}$$

where again $\vec{q}$ is written in terms of its parallel and perpendicular components and

$$|\vec{q} - \vec{k}_e|^2 = q^2 + k_e^2 - 2k_{e\perp}[k_{i\perp}\cos(\phi_{ki} - \phi_{ke}) - k_{f\perp}\cos(\phi_{kf} - \phi_{ke})] - 2k_{ez}q_z, \quad (15)$$

with $\phi_{ke} = \pi$ for the coplanar geometry used here.

*2.2 Partial Wave Amplitudes*

Because we are interested in OAM transfer from the projectile to the ionized electron, it is advantageous to evaluate Eq. (13) numerically. This allows for a straightforward insertion of a partial wave expansion for the ionized electron plane wave

$$\frac{e^{-i\vec{k}_e \cdot \vec{r}_2}}{(2\pi)^{3/2}} = \frac{1}{(2\pi)^{3/2}} \sum_{\lambda=0}^{\infty} (2\lambda + 1)(-i)^\lambda j_\lambda(k_e r_2) P_\lambda(\hat{k}_e \cdot \hat{r}_2), \quad (16)$$

where $j_\lambda(k_e r_2)$ is the spherical Bessel function, $P_\lambda(\hat{k}_e \cdot \hat{r}_2)$ is the Legendre polynomial, and $\lambda$ corresponds to the partial wave with angular momentum $\lambda\hbar$. Then, Eq. (13) becomes

$$T_{fi}^{PW} = \frac{4\pi}{(2\pi)^3 q^2 (2\pi)^{\frac{3}{2}}} \sum_{\lambda=0}^{\infty} (2\lambda + 1)(-i)^\lambda \int j_\lambda(k_e r_2) P_\lambda(\hat{k}_e \cdot \hat{r}_2)(-1 + e^{i\vec{q} \cdot \vec{r}_2}) \Phi(\vec{r}_2) d\vec{r}_2. \quad (17)$$

From this, the transition amplitudes for individual partial waves can be calculated, which show the relative importance of the amount of angular momentum transferred to the ionized electron. Insertion of Eq. (17) into the vortex transition amplitude of Eq. (9) then explicitly demonstrates the transfer of the quantized angular momentum of the incident vortex beam to the ionized electron.

### 3. Results

*3.1 Plane Wave vs. Electron Vortex Beam*

The use of EVBs as projectiles results in a larger kinematical parameter space than traditional plane wave (e,2e) collisions. In plane wave (e,2e) collisions, the kinematical parameters consist of the incident energy, ionized electron energy, and projectile scattering angle and therefore the FDCS are triply differential. For EVBs, there are also the parameters of opening angle $\alpha = \tan^{-1}\frac{k_{i\perp}}{k_{iz}}$ and incident OAM. Thus, the FDCS become 5-fold differential cross sections. In order to conduct a comprehensive study of the effect of these parameters, we explore a range of parameter space within acceptable limitations of the FBA and experimental limitations of EVB generation.

We begin by calculating the FDCS for incident plane waves and vortex beam electrons with energies of 500 eV and 1000 eV. Typical energies of experimental EVBs are on the order of a few hundred eV [8,9,11], and our use of incident energies of 500-1000 eV ensures that the perturbative FBA is valid. Specifically, the perturbation parameter $\eta = \left|\frac{Z_p}{v_p}\right|$ should be less than unity for the FBA to be valid, and the energies chosen here correspond to $\eta = 0.16$ (500 eV) and $\eta = 0.12$ (1000 eV). Three ejected electron energies of 20, 50, and 100 eV are used, again keeping in mind that the use of a plane wave to model the ejected electron requires larger energies. From (e,2e) studies, it is known that the differences between a plane wave treatment and other more sophisticated CCC, R-Matrix, and distorted wave treatments of ionized electrons are minimized for energies greater than 20 eV.

Typically, plane wave (e,2e) collisions involve scattering angles of a few degrees, depending on the incident projectile energies. We have chosen to use three scattering angles of 1, 10, and 100 mrad (0.0573, 0.573, and 5.73 degrees). Finally, we study the effect of the EVB parameters by examining projectiles with one or two units of OAM, as well as opening angles of 1, 10, and 100 mrad. We note that EVBs with angular momentum values up to a few hundred $\hbar$ have been produced [10], but for the sake of clarity we limit the scope of this work to small OAM values. Additionally, the results presented here indicate that the FDCS decrease in magnitude rapidly as OAM increases, making ionization by high OAM EVBs unlikely.

Figures 1-3 show the FDCS for incident plane wave and EVB projectiles with OAM of one for the range of kinematical values described above, and Figs 4-6 show the corresponding FDCS for EVBs with incident OAM of two. Although the basics of (e,2e) FDCS are well-known, it is worth mentioning some relevant trends observable in the plane wave FDCS in order to compare the FDCS for EVB projectiles. The plane wave FBA predicts a forward binary and backward recoil peak oriented along the plus and minus linear momentum transfer directions. As the scattering angle increases or the ionized electron energy decreases, the direction of the momentum transfer moves away from the beam direction. Also, as the ionized electron energy increases, the magnitude of the FDCS decreases. Analysis of the EVB results yields the following global trends. Like the plane wave FDCS, the EVB FDCS show a two peak structure. However, at the smallest scattering angle, there are no longer forward and backward peaks, but instead two forward peaks are located symmetrically to either side of the z-axis. At the largest scattering angle, the binary and recoil peak structure returns, but the peaks are no longer located at the linear momentum transfer direction. In fact, in most cases, a minimum is observed at or near the linear momentum transfer direction.

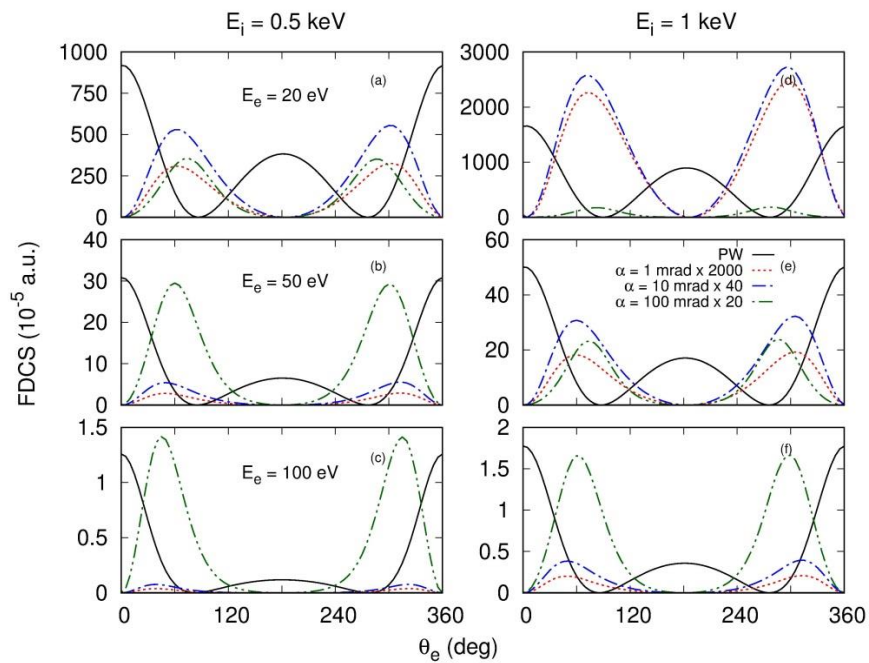

Figure 1 FDCS for ionization of hydrogen by plane wave (black line) and electron vortex beam projectiles with opening angles of 1 mrad (red dotted line), 10 mrad (blue dash dot line), and 100 mrad (green dash dot dot line). The EVB's OAM is $1\hbar$ and the projectile scattering angle is 1 mrad.

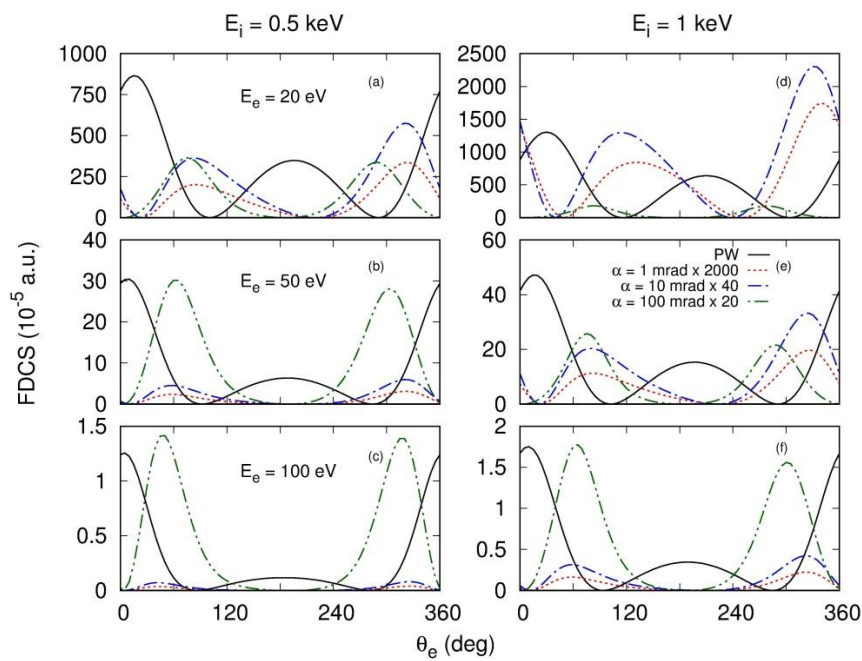

Figure 2 Same as Figure 1, but with a projectile scattering angle of 10 mrad.

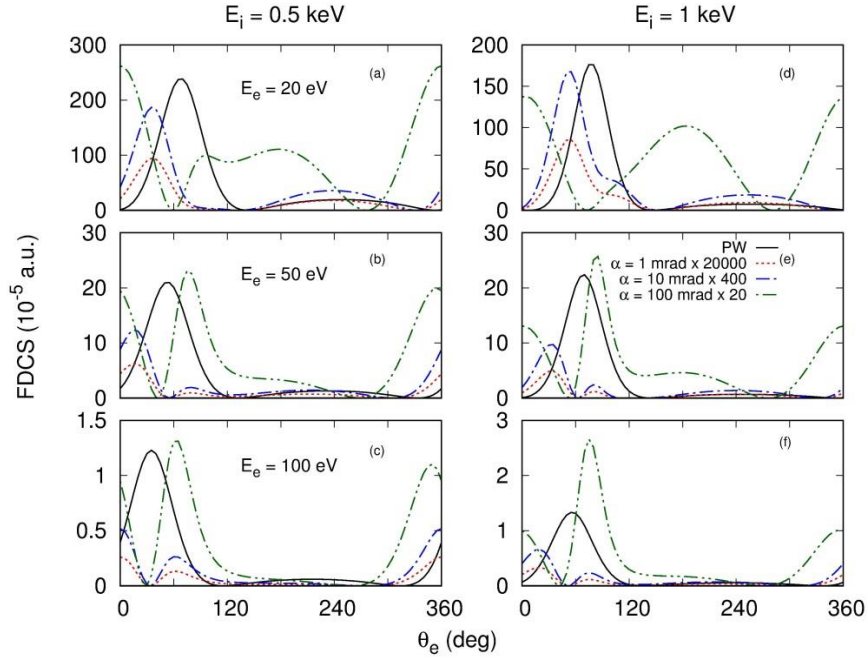

Figure 3 Same as Figure 1, but with a projectile scattering angle of 100 mrad.

Like the plane wave FDCS, the magnitude of the EVB FDCS decreases with increasing scattering angle and ionized electron energy. For the smallest scattering angle, the EVB peak magnitude and locations are nearly symmetric about $\theta_e = 180°$, but this symmetry is broken as the scattering angle increases. At the largest scattering angle, the binary peak is generally larger in magnitude and almost no recoil peak is observed. The two peaks that are present are both located in the forward direction on either side of the z-axis. While most of these trends hold for an incident OAM of two (Figs 4-6), the symmetry of peak location and magnitude does not exist in this case. Also, the overall magnitude of the EVB FDCS for OAM of two is 2-8 orders of magnitude smaller than the plane wave FDCS and 1-5 orders of magnitude smaller than the FDCS for OAM = 1. This indicates that ionization by EVB projectiles is most likely to occur for small quanta of OAM and large opening angle, and may only rarely occur for large angular momentum or small opening angle. While the FDCS for ionization by EVB is much smaller than the FDCS for ionization by plane wave, the EVB cross sections here are roughly the same order of magnitude as those for excitation of hydrogen by EVB [15]. This has important consequences for future experiments using EVBs because it indicates that any models for EVB applications will therefore need to include both excitation and ionization processes.

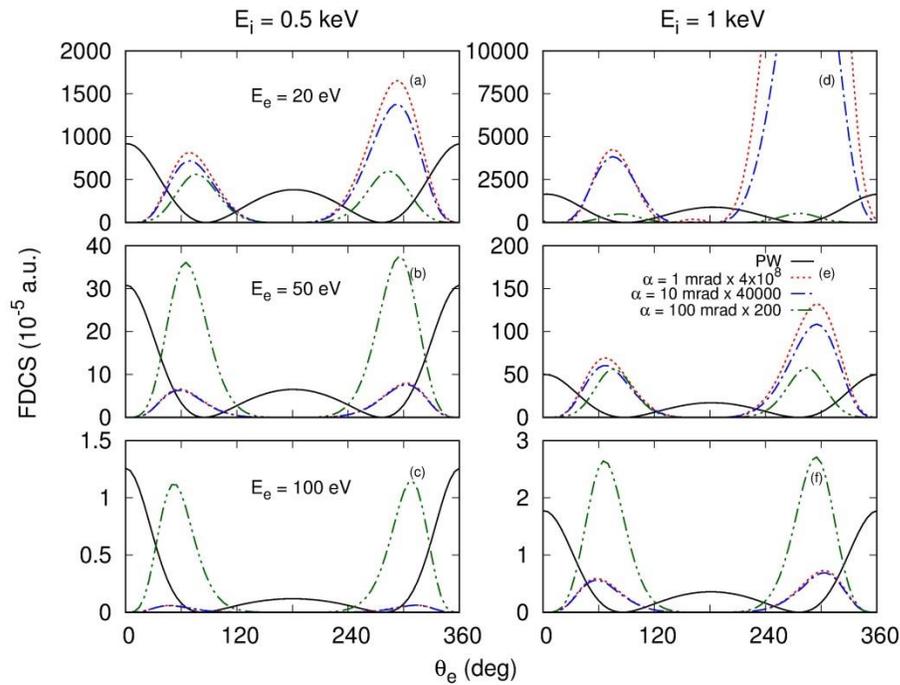

Figure 4 Same as Figure 1, but with an OAM = $2\hbar$.

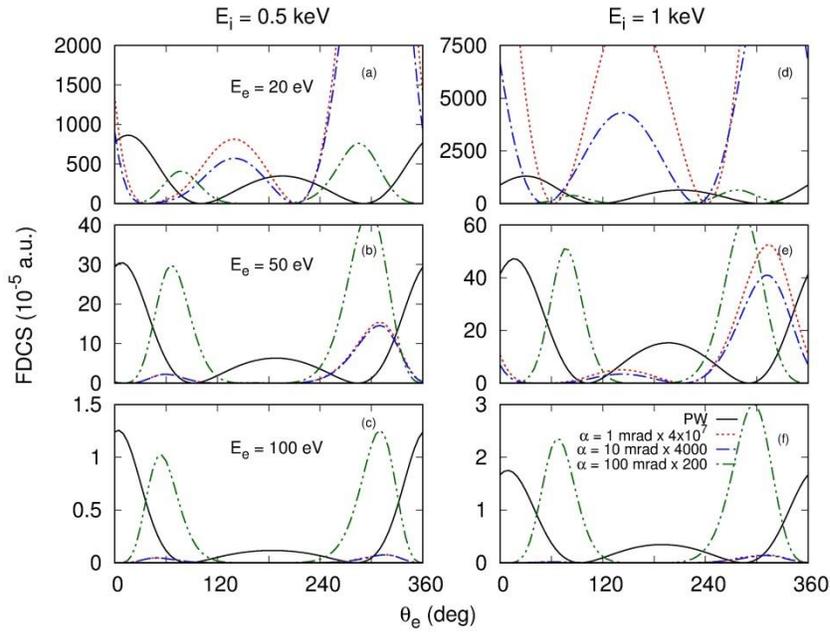

Figure 5 Same as Figure 2, but with an OAM = $2\hbar$.

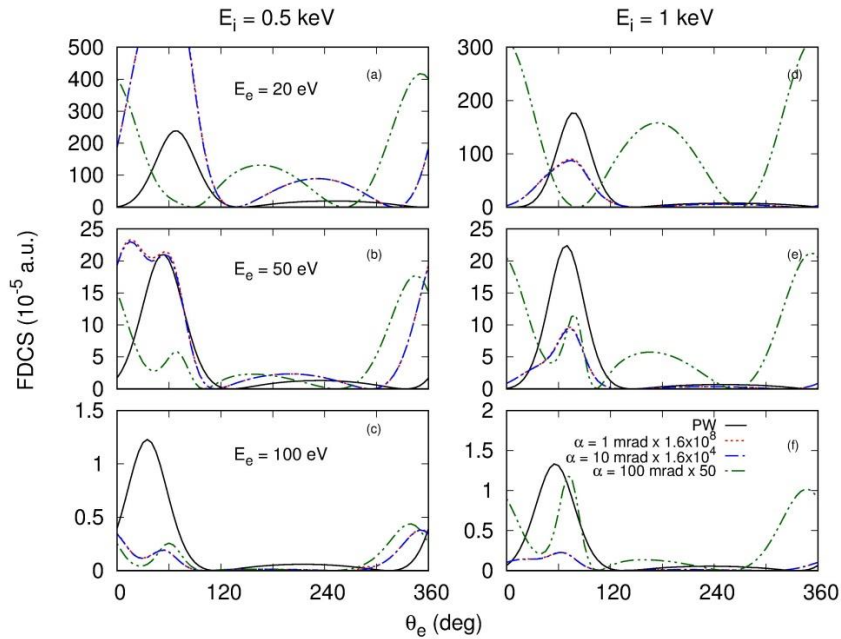

Figure 6 Same as Figure 3, but with an OAM = $2\hbar$.

*3.2 Angular Momentum Transfer*

Despite the small cross sections, one of the advantages of using EVB projectiles in (e,2e) collisions is the opportunity to study OAM transfer. Specifically, EVBs with discrete amounts of OAM combined with the partial wave amplitude analysis described in section 2.1 provide a unique opportunity to determine if and how the OAM is transferred to the ionized electron.

Figs 7 and 8 show the individual partial wave amplitudes and FDCS for plane wave and EVB projectiles with $E_i = 1\ keV$, $E_e = 20\ eV$, $\theta_s = 1\ mrad$, and OAM $= 1\hbar$ or $2\hbar$. Only the $\lambda = 1, 2$, and $3$ amplitudes are shown because these are the three that contribute significantly to the shape and magnitude of the total amplitude. The $\lambda = 0$ amplitudes are constant with the values listed in Table 1. These kinematics were chosen for more detailed study because they have the largest FDCS of the kinematical parameters used here.

|  | Plane Wave | EVB OAM $= 1\hbar$ $\alpha = 1$ mrad | EVB OAM $= 1\hbar$ $\alpha = 10$ mrad | EVB OAM $= 2\hbar$ $\alpha = 1$ mrad | EVB OAM $= 2\hbar$ $\alpha = 10$ mrad |
|---|---|---|---|---|---|
| FDCS | 1.5 x 10$^{-6}$ | 6.9 x 10$^{-15}$ | 7.3 x 10$^{-13}$ | 3.8 x 10$^{-24}$ | 3.8 x 10$^{-20}$ |
| $Re(T_{fi})$ | 1.81 x 10$^{-3}$ | -2.0 x 10$^{-17}$ | -1.5 x 10$^{-16}$ | 1.8 x 10$^{-12}$ | 1.8 x 10$^{-10}$ |
| $Im(T_{fi})$ | 4.4 x 10$^{-17}$ | -7.6 x 10$^{-8}$ | -7.6 x 10$^{-7}$ | -4.5 x 10$^{-19}$ | 3.6 x 10$^{-18}$ |

Table 1 Ejected electron partial wave FDCS and amplitudes in a.u. for $\lambda = 0$. The incident projectile energy is 1 keV, the ionized electron energy is 20 eV, and the scattering angle is 1 mrad.

The partial wave analysis shows some interesting features and we again begin with some observations of the plane wave amplitudes. From Fig. 7 and Eq. (14), it is obvious that the plane wave amplitude is purely real and the magnitude of the individual partial wave amplitudes decreases with increasing $\lambda$. For a plane wave projectile, the largest amplitude is for $\lambda = 1$, indicating that the ionized electron is most likely to leave the collision with one unit of angular momentum. However, the $\lambda = 1$ FDCS shows equal magnitude binary and recoil peaks and the correct magnitudes are only obtained when the $\lambda = 2$ amplitude is included. The binary peak amplitudes constructively interfere, while the recoil peak amplitudes destructively interfere to give the resulting large binary and small recoil peaks in the FDCS. The number of peaks in the plane wave amplitudes and FDCS correspond to the appropriate shapes for s, p, d, f, etc waves, and all plane wave partial wave amplitudes are symmetric about the momentum transfer direction.

For EVB projectiles with OAM = 1, the partial wave amplitudes are purely imaginary, while for OAM = 2, they are purely real. This is due to the $(-i)^l$ factor in the incident Bessel beam expression of Eq. (5). For both values of incident OAM, the EVB partial wave amplitudes are smaller than their plane wave counterparts, which is expected from the relative magnitudes of the FDCS. Unlike the plane wave partial wave amplitudes, which are symmetric about $\theta_e = 180°$, the EVB amplitudes for OAM = 1 are antisymmetric about $\theta_e = 180°$ with nodes at $\theta_e = 0°$ and $180°$ and they sum to produce a total amplitude that is antisymmetric about $\theta_e = 180°$. The EVB amplitudes for OAM = 2 are symmetric for $\lambda > 1$, but antisymmetric for $\lambda = 1$. This difference in symmetries results in significant shape differences in the EVB FDCS are observed for OAM = 1 compared with OAM = 2. Specifically, the two peaks in the FDCS for OAM = 1 are located symmetrically about the momentum transfer direction and with equal magnitude. The FDCS for OAM = 2 on the other hand have a much larger recoil peak than binary peak. This can be traced directly to interference of the individual partial wave amplitudes.

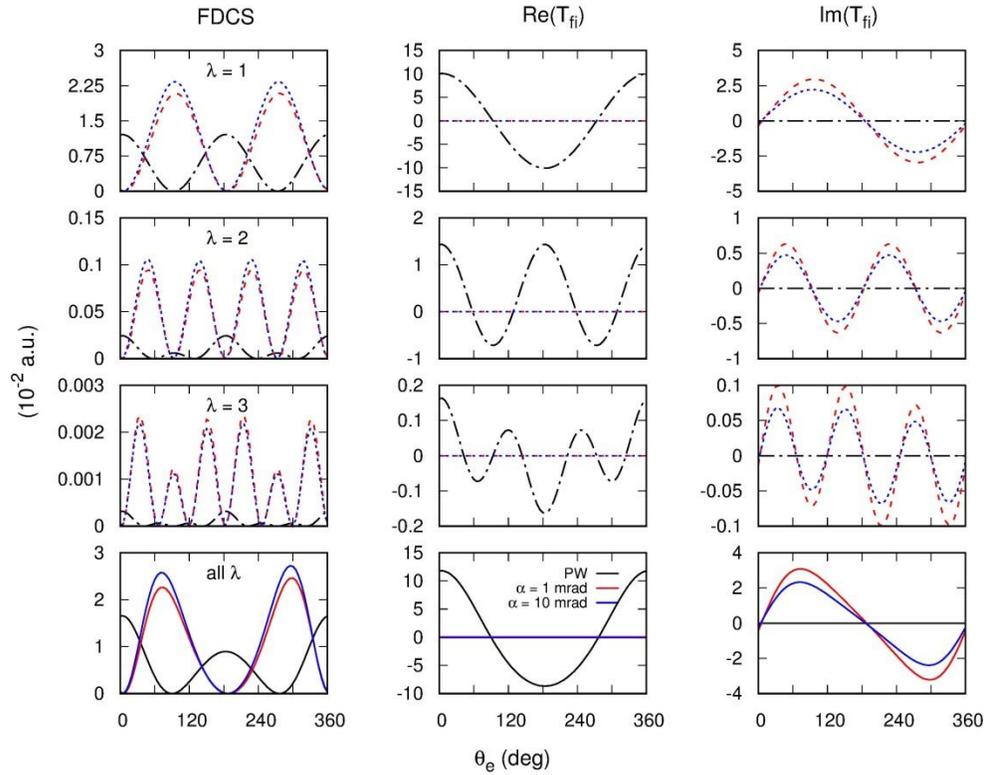

Figure 7 FDCS and transition amplitudes for ionization of hydrogen by plane wave (black line) and electron vortex beam projectiles with opening angles of 1 mrad (red lines) and 10 mrad (blue lines). Results are shown for individual partial waves of the ionized electron (rows 1-3) and all partial waves (row 4). The incident projectile energy is 1 keV, the ionized electron energy is 20 eV, the scattering angle is 1 mrad, and the EVB's OAM is $1\hbar$. In column 1, the FDCS for EVB ionization have been multiplied by 2000 ($\alpha = 1$ mrad) and 40 ($\alpha = 10$ mrad); in columns 2 and 3, the amplitudes for $\alpha = 1$ mrad have been multiplied by 10. The plane wave amplitudes and FDCS are absolute. The FDCS and amplitudes for $\lambda = 0$ are constant with respect to ionized electron angle and their values are listed in Table 1.

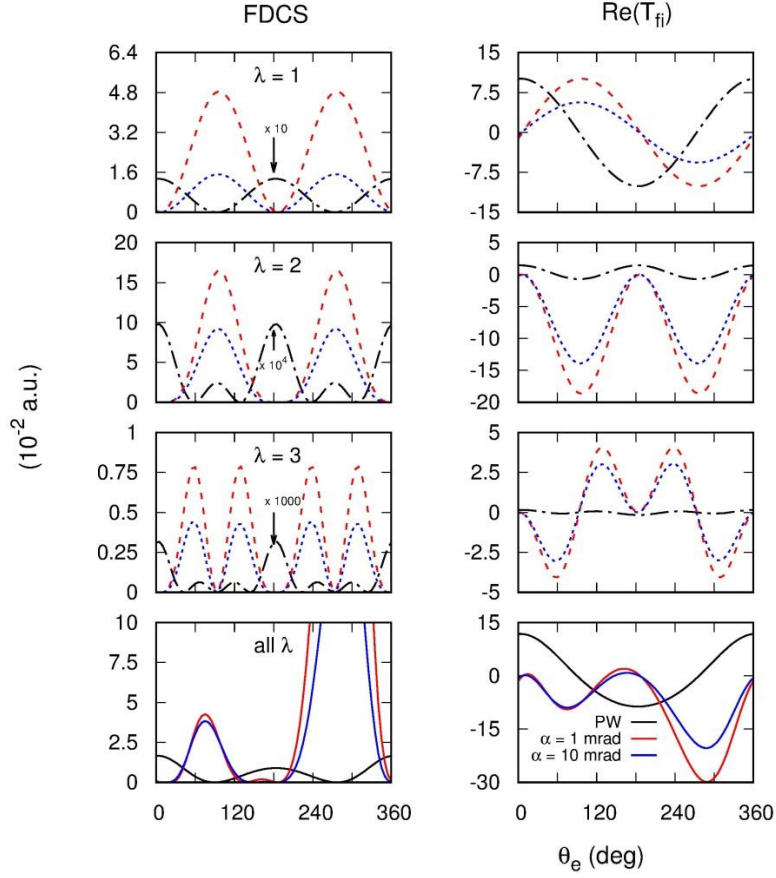

Figure 8 Same as Figure 7, but with OAM = $2\hbar$. In column 1, the FDCS for EVB ionization have been multiplied by 4 x $10^8$ ($\alpha = 1$ mrad) and 4000 ($\alpha = 10$ mrad); in column 2, the amplitudes have been multiplied by $10^5$ ($\alpha = 1$ mrad) and 100 ($\alpha = 10$ mrad). The plane wave amplitudes and FDCS are absolute. The FDCS and amplitudes for $\lambda = 0$ are constant with respect to ionized electron angle and their values are listed in Table 1.

Comparison of the EVB amplitudes for OAM = 1 and 2 shows that for OAM = 1, the $\lambda = 1$ amplitude is the largest, while for OAM = 2, the $\lambda = 2$ amplitude is the largest. This indicates that if ionization occurs, the incident electron likely transfers all of its angular momentum to the ionized electron.

## 4. Conclusion

We have presented fully differential cross sections for ionization of hydrogen by electron vortex beam projectiles calculated within the first Born approximation. A range of kinematical parameters were used in order to study the effect of incident energy, ionized electron energy, and scattering angle. In general, the FDCS for the EVB projectiles are several orders of magnitude smaller than those of the plane wave projectiles. The shape of the FDCS for EVBs is also significantly altered from those of the plane wave projectile. In the case of plane waves, the FBA predicts binary and recoil peaks oriented along and opposite of the linear momentum transfer direction. For EVBs, no single linear momentum transfer direction can be defined, and the binary and recoil peaks in the FDCS are shifted from the plane wave predicted locations, with a minimum typically occurring near the linear momentum transfer direction.

Unlike their plane wave counterparts, EVBs carry discrete quantities of OAM and are characterized by their opening angle which relates transverse and longitudinal momenta. This OAM can be transferred to other particles during the collision process. By expanding the ionized electron wave function in terms of partial waves, we were able to determine that the OAM of the incident EVB is transferred directly to the ionized electron. The partial wave amplitudes also showed zero amplitude for ejected electrons to be found near the linear momentum transfer direction.

While the FBA is a simplistic model for (e,2e) collisions, the results presented here are to our knowledge the first FDCS for ionization by EVB. The theoretical derivation for the EVB transition matrix can be written in terms of the plane wave transition matrix, making expansion to more sophisticated models and more complicated targets straightforward. We are currently expanding the FBA model to include a distorted wave treatment of the ionized electron so that slower ejected electron energies can be studied. In addition, we are also generalizing the model to multi-electron targets.

## Acknowledgements

We gratefully acknowledge the support of the NSF under Grant No. PHY-1505217.